# MODELING TURBULENT MIXING AND SAND DISTRIBUTION IN THE BOTTOM BOUNDARY LAYER

Rafik Absi[1]

[1] EBI, 32 Bd du Port, 95094 Cergy-Pontoise cedex, France.
E-Mail : r.absi@ebi-edu.com .

**Abstract**: For the calculation of turbulent mixing in the bottom boundary layer, we present simple analytical tools for the mixing velocity $w_m$ and the mixing length $l_m$. Based on observations of turbulence intensity measurements, the mixing velocity $w_m$ is represented by an exponential function decaying with $z$. We suggest two theoretical functions for the mixing length, a first $l_{m1}$ obtained from the $k$-equation written as a constant modeled fluctuating kinetic energy flux and a second $l_{m2}$ based on von Kármán's similarity hypothesis. These analytical tools were used in the finite-mixing-length model of Nielsen and Teakle (2004). The modeling of time-mean sediment concentration profiles $C(z)$ over wave ripples shows that at the opposite of the second equation $l_{m2}$ which increases the upward convexity of $C(z)$, the first equation $l_{m1}$ increases the upward concavity of $C(z)$ and is able to reproduce the shape of the measured concentrations for coarse sand.

**INTRODUCTION**

The prediction of coastal sediment transport depends mainly on the turbulence model used in the bottom boundary layer. The mechanism of turbulence is of an extremely complicated nature. In the turbulent bottom boundary layer, the turbulent mixing motion is responsible for an exchange of momentum, and it enhances the transfer of mass. Under waves, the turbulent mixing which generates a net vertical flux of suspended sediment can be modeled using realistic turbulence parameters together with simple analytical methods. Indeed, the development of theoretical and semi-theoretical analytical methods for turbulent flows is of great importance in both practical engineering applications and basic turbulence research.

In this paper, we will present simple analytical tools for the calculation of turbulent mixing. The turbulent mixing can be described simply by a mixing velocity $w_m$ and a mixing length $l_m$. We will first present an analytical expression for the mixing velocity (Absi 2000, Nielsen and Teakle 2004, and Absi 2005) which is based on observations of turbulence intensity measurements and confirmed theoretically. We will present, on the other hand, two theoretical algebraic equations for the mixing length, a first equation obtained from the turbulent kinetic energy *k*-equation (Nezu and Nakagawa 1993, Absi 2005) with some basic assumptions (steady flow, local equilibrium and the proposed equation for the mixing velocity), and a second equation based on the similarity hypothesis (Absi 2002) with some assumptions (local equilibrium and the proposed equation for the mixing velocity). We will finally apply the proposed analytical methods to the finite-mixing-length model of Nielsen and Teakle (2004) for the modeling of time-mean suspended sediment concentration profiles in an oscillatory boundary layer over wave ripples. The numerical solutions will be compared with measurements from McFetridge and Nielsen (1985).

**THE MIXING VELOCITY**

The mixing velocity $w_m$ profile could be obtained from observations of turbulence intensity measurements. From turbulence measurements of Wijetunge and Sleath (1998) (figure 1), we noticed (Absi, 2000) that turbulence intensity $(\overline{u'^2})^{1/2}$ decreases exponentially with *z* for mobile beds (plane and with ripples) and therefore can be expressed by

$$\frac{(\overline{u'^2})^{1/2}}{U_0} \approx \exp(-C_1 \xi) \qquad (1)$$

where $\xi = z/h$, $h$ = the scale of the flow can represent either the boundary layer thickness or the turbulent flow depth (m), $U_0$ = the maximum value of the free stream velocity (m/s), and $C_1$ = constant. Or as

$$(\overline{u'^2})^{1/2} \approx U_* \exp(-C_1 \xi) \qquad (2)$$

where $U_*$ = the friction velocity (m/s). This expression is valid only for mobile beds (plane and with ripples). We can write therefore the mixing velocity in the form

$$w_m = \gamma U_* \exp(-C_1 \xi) \qquad (3)$$

where $\gamma$ = constant. Or in the form $w_m = w_m(z_0) \exp(-z/L_w)$ where $w_m(z_0) = \gamma U_*$, $z_0$ = the hydrodynamic roughness (m) and $L_w = h/C_1$ (Nielsen and Teakle 2004). If we assume that $v' = C_v u'$ and $w' = C_w u'$, with $C_v$ and $C_w$ are constants in the energy equilibrium region (see Nezu and Nakagawa 1993), we can write $\sqrt{k} \approx \left(\overline{u'^2}\right)^{\frac{1}{2}}$ and therefore

$$\sqrt{k} \approx U_* \exp(-C_1 \xi), \tag{4}$$

where $k$ = the turbulent kinetic energy (m²/s²).

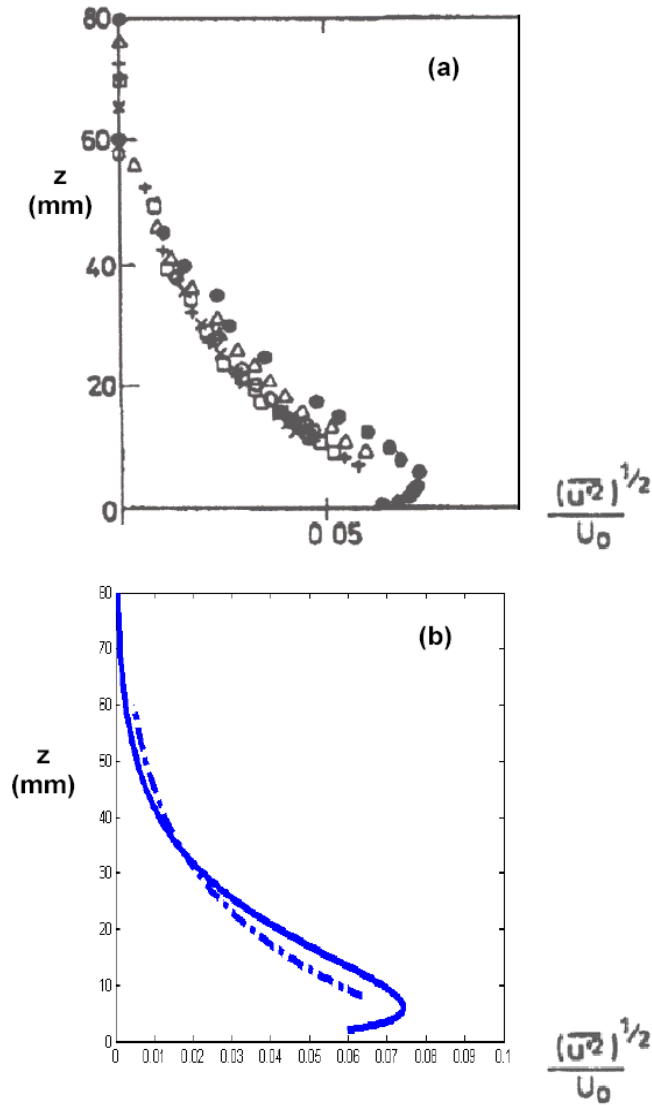

Fig. 1. Vertical distribution of dimensionless turbulent intensity: (a) Measurements of Wijetunge and Sleath (1998) for fixed bed: '•'; and mobile beds: '∆' '+' ' ' '×' 'o'; (b) The dash-dot line given by equation (1) reproduces mobile beds measurements (while the solid line reproduces fixed bed measurements, see Absi 2006).

On the other hand, equation (4) represents a solution of the modeled *k*-equation when the flow is steady and in local equilibrium. Indeed, in local equilibrium region, where the energy production is balanced by the dissipation, Nezu and Nakagawa (1993) wrote for steady open-channel flows the modeled *k*-equation as a constant modeled fluctuating kinetic energy flux

$$\nu_t \frac{dk}{dz} = const \qquad (5)$$

where $\nu_t$ = the eddy viscosity (m²/s). They wrote an approximation for $\nu_t$ as

$$\nu_t \approx \frac{k^2}{\varepsilon} \approx \frac{l}{k}\left(\frac{k}{u'}\right)^3 \approx \left(\frac{U_*^2}{k}\right)(U_* h) \approx \frac{1}{k} \qquad (6)$$

where $\varepsilon$ = the energy dissipation (m²/s³) and $l$ = a length scale. For local equilibrium, an explicit relation between the length scale $l$ and the mixing length $l_m$ is given by $l_m = (C_\mu)^{-\frac{1}{4}} l$; where $C_\mu$ = the empirical constant in the $k-\varepsilon$ model. By replacing (6) into (5)

$$\left(\frac{1}{k}\right)\left(\frac{dk}{d\xi}\right) = const \equiv -2C_1, \qquad (7)$$

and by integrating (7), Nezu and Nakagawa proposed a semi-theoretical function for *k*

$$\frac{k}{U_*^2} = D \exp(-2C_1 \xi) \qquad (8)$$

where $D$ = a constant. We can write (8) in the form

$$\sqrt{k} = \sqrt{k_0} \exp[-C_1 (\xi - \xi_0)], \qquad (9)$$

where $k_0 = k(z_0)$.

**THE MIXING LENGTH**

Near the bottom, the turbulent length scale is estimated to be proportional to the size of the large eddies, those that contain the most energy, and thus the most momentum. From a certain distance *z*, if we assume that the most effective eddies for the mixing are precisely those of size *z*, we can write $l \sim z$ which gives the Prandtl mixing length equation

$$l_m = \kappa z \qquad (10)$$

where $\kappa$ = 0.4 (the von Kármán constant). In a turbulent boundary layer, the largest eddies are limited by the transverse dimension of the flow namely the boundary layer thickness, we are able therefore to write

$$l_m \sim h \qquad (11)$$

The linear mixing length profile seems not realistic, because physically the mixing length cannot increase linearly over the entire boundary layer or flow depth. We will suggest two theoretical equations for $l_m$ based on $w_m$.

**A first mixing length equation**
*Weakness of Nezu and Nakagawa's demonstration*
Even if the shape for $k$ is realistic and seems correspond to experimental measurements, the demonstration of Nezu and Nakagawa presents a weakness and does not allow justifying this form theoretically. Indeed, the solution (8) allows to write (12) and not (7).

$$\left(\frac{\exp(-3C_k \xi)}{k}\right)\left(\frac{dk}{d\xi}\right) = const \equiv -2C_1 \qquad (12)$$

Equation (12) shows therefore a contradiction in the demonstration of Nezu and Nakagawa.

*Proposed demonstration*
If we assume a shape $f(\xi)$ as $k/U_*^2 = D\, f(\xi)^2$ and $u'/U_* = D_u\, f(\xi)$, a more coherent approximation is (Absi 2005)

$$\nu_t \approx \frac{k^2}{\varepsilon} \approx \frac{l}{k}\left(\frac{k}{u'}\right)^3 \approx \frac{l(\xi)}{k} U_*^3 f(\xi)^3 \qquad (13)$$

We can therefore write (7) as

$$\left(\frac{g(\xi)}{\sqrt{k}}\right)\left(\frac{d\sqrt{k}}{d\xi}\right) = const \equiv -C_n \qquad (14)$$

where $g(\xi) = l(\xi)\, f(\xi)^3$ and $C_n$ = a constant. By integrating (14) between $\xi_0 = z_0/h$ and $\xi = z/h$, we obtain

$$\sqrt{k} = \sqrt{k_0} \exp\left(-C_n \int_{\xi_0}^{\xi} \frac{d\xi}{g(\xi)}\right) \qquad (15)$$

From this equation, we find equation (8) only in the case where $g(\xi) = l(\xi)\, f(\xi)^3 = const$, and this is possible only for one form of $l$. With $f(\xi) = \exp(-C_1 \xi)$ (here $C_1 = C_n/const$), the sole expression possible for $l$ is given by (Absi 2005)

$$l(\xi) = const\, \exp(3C_1 \xi) \qquad (16)$$

From this equation and (8), we have $g(\xi)=const$ and therefore (8) is a solution for (5). If we take a boundary condition for $l_m$ at $z_0$, $(l_m)_0 = l_m(z_0) = \kappa z_0$, we can write for rough mobile beds, a mixing length equation as $l_m(\xi) = \kappa h \xi_0 \exp(3 C_1 (\xi - \xi_0))$ or

$$l_m(z) = \kappa z_0 \exp\left(3 \frac{z-z_0}{L_w}\right) \qquad (17)$$

where $h = C_1 L_w$. We refer to this first mixing length profile as $l_{m1}$. This vertical profile is concave downward (figure 2). We notice that, the slope at the origin $z_0$, equal to $(3 \lambda z_0)/L_w$, increases with roughness $z_0$. In figure 2, we have $\lambda$ instead of $\kappa$=0.4. We wrote $(l_m - \lambda z_0)/\lambda(L_w - z_0)$ in order to represent a dimensionless mixing length profile which is valid for any value of $\lambda$.

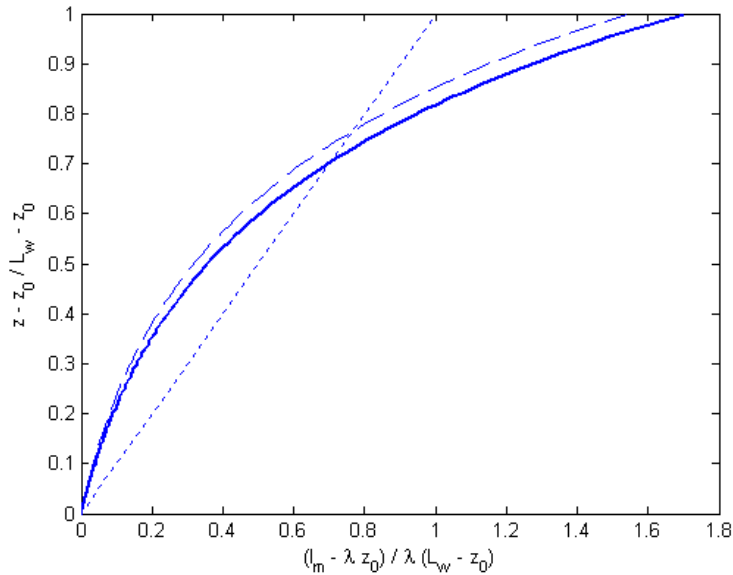

Fig. 2. Mixing length profile $l_{m1}(z) = \lambda z_0 \exp\left(3\frac{z-z_0}{L_w}\right)$, solid line: $l_{m1}$ for $z_0/L_w$=0.1136 ($L_w$=0.044m), dashed line $l_{m1}$ for $z_0/L_w$=0.1 ($L_w$=0.05m), dotted line: $l_m = \lambda z$.

**A second mixing length equation based on an extension of von Kármán similarity hypothesis**

On the other hand, the von Kármán's similarity hypothesis, which assumes that turbulent fluctuations are similar at all points of the field of flow (similarity rule), gives the mixing length in the form

$$l_m = -\kappa \left[ \frac{\left(\frac{\partial u}{\partial z}\right)}{\left(\frac{\partial^2 u}{\partial z^2}\right)} \right] \qquad (18)$$

For local equilibrium, the energy production is balanced by the dissipation. This allows writing $\partial u/\partial z \sim \sqrt{k}/l_m$, where $k$ is the turbulent kinetic energy. An extension of von Kármán's similarity hypothesis allows us therefore to write

$$l_m = -\kappa \left[ \frac{\frac{\sqrt{k}}{l_m}}{\frac{\partial}{\partial z}\left(\frac{\sqrt{k}}{l_m}\right)} \right] \qquad (19)$$

We can write (19) in the form $-\frac{f'}{f^2} = \frac{\kappa}{\sqrt{k}}$, with $f = \frac{\sqrt{k}}{l_m}$; applying the condition $l_m(z=z_0) = \kappa z_0$ and by integrating $\int_{z_0}^{z} -\frac{f'}{f^2} dz = \frac{1}{f} - \frac{1}{f_0} = \kappa \int_{z_0}^{z} \frac{1}{\sqrt{k}} dz$, we obtain

$$l_m = \kappa \sqrt{k} \left[ \int_{z_0}^{z} \frac{1}{\sqrt{k}} dz + \frac{z_0}{\sqrt{k_0}} \right] \qquad (20)$$

Equation (20) can be integrated if the turbulent kinetic energy is given by an algebraic equation. We write (8) as

$$\sqrt{k(z)} = \sqrt{D} \, U_* \exp\left(-\frac{z}{L_w}\right) \qquad (21)$$

Inserting (21) into (20) and by integrating (Absi, 2002), we obtain

$$l_m(z) = \kappa e^{(-z/L_w)} \left[ L_w e^{(z/L_w)} - L_w e^{(z_0/L_w)} + z_0 e^{(z_0/L_w)} \right] \qquad (22)$$

This equation is in the form $l_m \sim L_w$. In fact, when the two last terms in (22) becomes smaller than the first term, we find $l_m = \kappa L_w$. This equation, which confirms the hypothesis (11), shows that an increase in the boundary layer thickness implies an increase in the mixing length. We can write this mixing length equation (22) in the form

$$l_m(z) = \kappa \cdot \left[ L_w - (L_w - z_0) \cdot e^{\left(-\frac{(z-z_0)}{L_w}\right)} \right] \quad (23)$$

We refer to this second mixing length profile as $l_{m2}$. For a smooth wall ($z_0 = 0$), we write equation (23) as

$$\frac{l_m(z)}{L_w} = \kappa \cdot \left[ 1 - e^{\left(-\frac{z}{L_w}\right)} \right] \quad (24)$$

This mixing length profile, which is non-linear, is different from Prandtl's profile. From a same (imposed) value $\kappa \cdot z_0$ in $z_0$, our mixing length (equation 23) increases more slowly with z, the gradient $dl/dz$, which is equal to $\kappa \cdot (1 - z_0/L_w) \cdot \exp(-(z-z_0)/L_w)$, is everywhere (since $z \geq z_0$) smaller than $\kappa$. Moreover, the slope at the origin $z_0$, equal to $\kappa \cdot (1 - z_0/L_w)$, decreases with the roughness $z_0$ (figure 3); this involves the introduction of an effective kappa $\kappa^*$ as

$$\kappa^* = \kappa \cdot (1 - z_0/L_w) \quad (25)$$

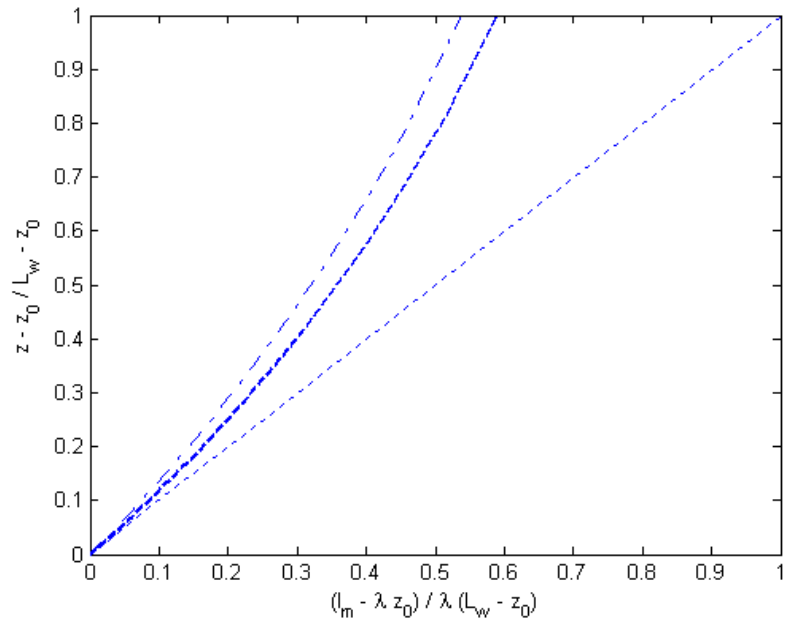

Fig. 3. Mixing length profile $l_{m2}(z) = \lambda \left[ L_w - (L_w - z_0) \exp\left(-\frac{(z-z_0)}{L_w}\right) \right]$, dashed line: $l_{m2}$ for $z_0/L_w = 0.1136$ ($L_w = 0.044$m), dash-dot line: $l_{m2}$ for $z_0/L_w = 0.2272$ ($L_w = 0.022$m), dotted line: $l_m = \lambda z$.

The difference $\kappa^* \neq \kappa$ shows that even at the first order (linearization on the parameter $(z-z_0)/L_w$ which is assumed to be small, near the bottom), our profile is different from the Prandtl's profile $\kappa \cdot z$. It is written as:

$$l_m(z) \approx \kappa^* \cdot z + \kappa \cdot \frac{z_0^2}{L_w} \qquad (26)$$

The two profiles are different even near the bottom; they will be identical in the sole case of low roughness (smooth bottom). Only here, $\kappa^* \approx \kappa$ and $l_m(z) \approx \kappa \cdot z$. We can conclude that the equation $\kappa \cdot z$ is able to describe the vertical profile of mixing length near the bottom and only for smooth bottom. As for figure 2, in figure 3 we have $\lambda$ instead of $\kappa = 0.4$. We wrote $(l_m - \lambda z_0) / \lambda(L_w - z_0)$ in order to represent a dimensionless mixing length profile which is valid for any value of $\lambda$.

## SUSPENDED SEDIMENT PROFILES OVER WAVE RIPPLES
### The Finite-mixing-length model

The main idea of the finite-mixing-length model (Nielsen and Teakle 2004) is that the Fickian diffusion is not the right theoretical framework for sediment suspension. The swapping of fluid parcels (including suspended sediment) between different levels can generate a net vertical flux. If the parcels travel vertically with equal and opposite velocities, the resulting sediment flux density is (see Teakle and Nielsen 2004)

$$q_m = w_m \cdot [c(z - l_m/2) - c(z + l_m/2)] \qquad (27)$$

and by Taylor expansion

$$q_m = -w_m l_m \left[ \frac{dc}{dz} + \frac{l_m^2}{24} \frac{d^3 c}{dz^3} + \cdots \right] \qquad (28)$$

This theory contains Fickian diffusion as the limiting case since the Fickian approximation is obtained by retaining only the first term in the [ ].

Time averaged concentrations of suspended sediment result from the balance between an upward mixing flux and a downward settling flux

$$q_m - w_s c(z) = 0 \qquad (29)$$

With (28), equation (29) becomes

$$-w_m l_m \left[ \frac{dc}{dz} + \frac{l_m^2}{24} \frac{d^3 c}{dz^3} + \cdots \right] - w_s c(z) = 0 \tag{30}$$

By including only the first two terms of the Taylor expansion, we obtain

$$\frac{l_m^2}{24} \frac{d^3 c}{dz^3} + \frac{dc}{dz} + \frac{w_s}{w_m l_m} c(z) = 0 \tag{31}$$

The third-order Ordinary Differential Equation (31), which approximates the finite-mixing-length model, needs vertical profiles for mixing velocity $w_m$ and mixing length $l_m$. The simulations of Nielsen and Teakle presented in figure (4) are obtained with a mixing velocity profile given by $w_m = w_m(z_0) \exp(-z/L_w)$, and a mixing length as $l_m = \lambda z$ (with $\lambda = 1$). In figure (4), the numerical solutions (Teakle and Nielsen 2004) are compared with measurements from McFetridge and Nielsen (1985). These simulations show that the finite-mixing-length solution is similar to the Fickian solution for fine sand, whereas the finite-mixing-length solution demonstrates enhanced mixing efficiency compared to the corresponding Fickian solution for coarse sand.

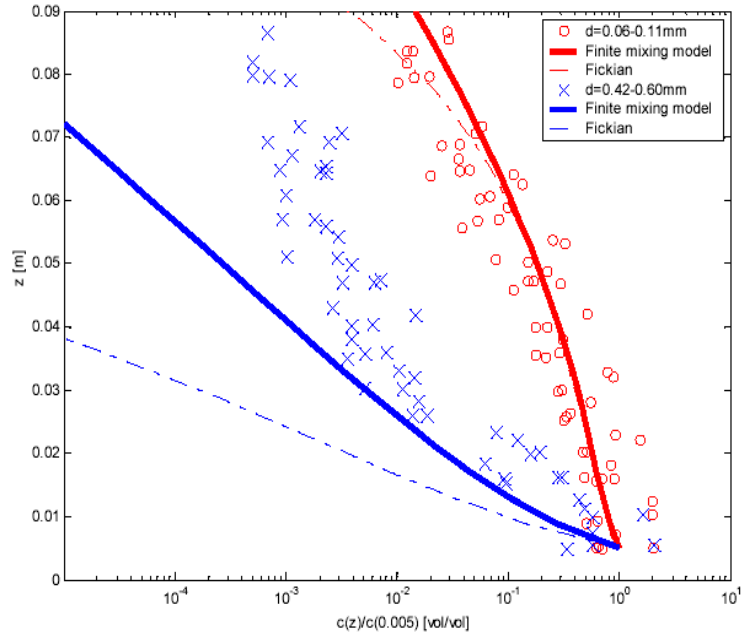

Fig. 4. Concentration profiles over wave ripples. O: Fine sand, $w_s$=0.65cm/s, X: coarse sand, $w_s$=6.1cm/s (measurements from McFetridge and Nielsen, 1985). Solid lines: The numerical solution to the third-order ODE (31) with $l_m = \lambda z$ ($\lambda$=1) and $z_0$=0.005m, $w_m(z_0)$=0.025m/s, $L_w$=0.022m. Dash-dot lines show the Fickian approximations, corresponding to dropping the first term in (31). (Figure: Teakle and Nielsen, 2004).

The upward convex profile for fine sand versus upward concave for the coarse sand is reproduced by the model (Nielsen and Teakle 2004). Even if the upward concave concentration profile for coarse sand (figure 4) is reproduced with $l_m = \lambda\, z$, there is a difference with experimental data (McFetridge and Nielsen 1985) at the top ($z$ between 0.04 and 0.09m) which seems to become more important for $z>0.09$m. This imperfection could be related to the linear mixing length profile. In fact, the linear mixing length profile seems not realistic, because physically the mixing length cannot increase linearly over the entire boundary layer or flow depth. In figure 4, the parameters $w_m(z_0)=0.025$m/s and $L_w=0.022$m correspond better to a fit for fine sand. The solution is perfect for fine sand but under-estimates the coarse sand concentrations. A more appropriate fit, which represents a better compromise, can correspond to a value of $L_w$, which is approximately equal to twice the Nielsen and Teakle's value. In order to evaluate the proposed $l_m$ equations, we will apply them in the modeling of period-averaged sediment concentration profiles.

**Results and discussion**

We will study the influence of the two proposed mixing length $l_m$ equations on the shape of concentration profiles $C(z)$. Our main question is: Are the proposed $l_m$ profiles capable of changing the shape of $C(z)$ and improving the computed $C(z)$?

Even if $l_{m1}$ has been established for steady flows, it seems to be a good tool for the computation of time-mean concentration profiles.

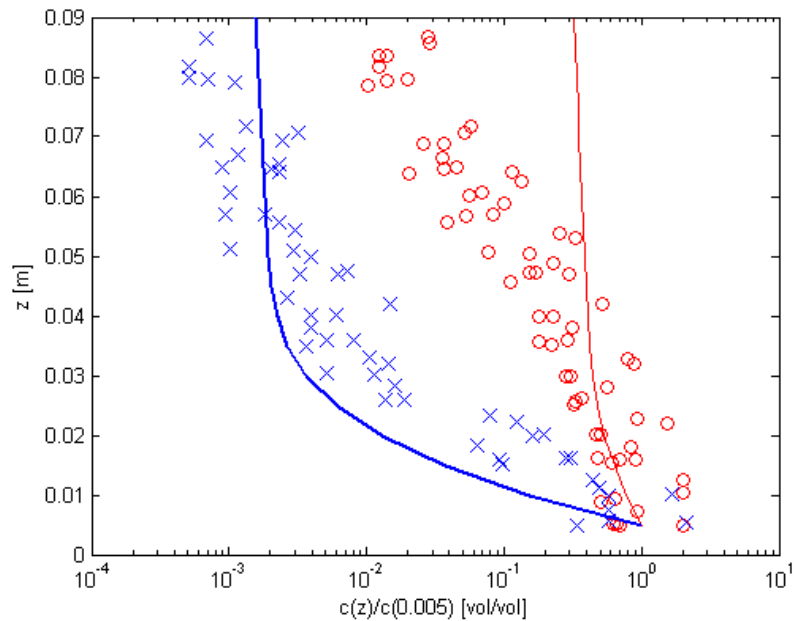

Fig. 5. Time-averaged concentration profiles over wave ripples. Curves: numerical solution to the third-order ODE (31) with $l_{m1}(z) = \lambda\, z_0 \exp\left(3\,\dfrac{z-z_0}{L_w}\right)$ ; ($\lambda=1$) , $L_w=0.044$m, $z_0=0.005$m; $w_m(z_0)=0.025$m/s for fine and coarse (bold line) sand. Measurements: O Fine sand ($w_s=0.65$cm/s); and X coarse sand, ($w_s=6.1$cm/s).

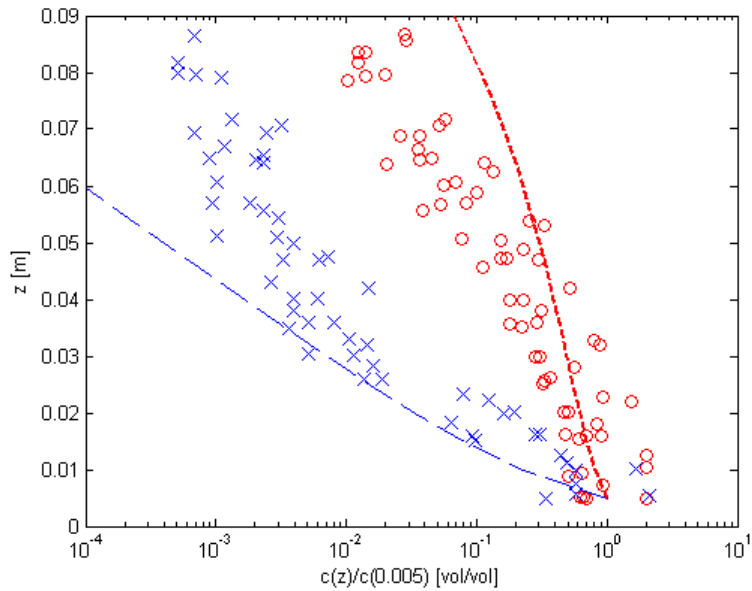

Fig. 6. Time-averaged concentration profiles over wave ripples. Curves: numerical solution to the third-order ODE (31) with $l_{m2}(z)=\lambda\left[L_w-(L_w-z_0)\exp\left(-\frac{(z-z_0)}{L_w}\right)\right]$, ($\lambda=1$) ; $L_w$=0.044m ; for fine (bold) and coarse sand. Measurements: O Fine sand and X coarse sand.

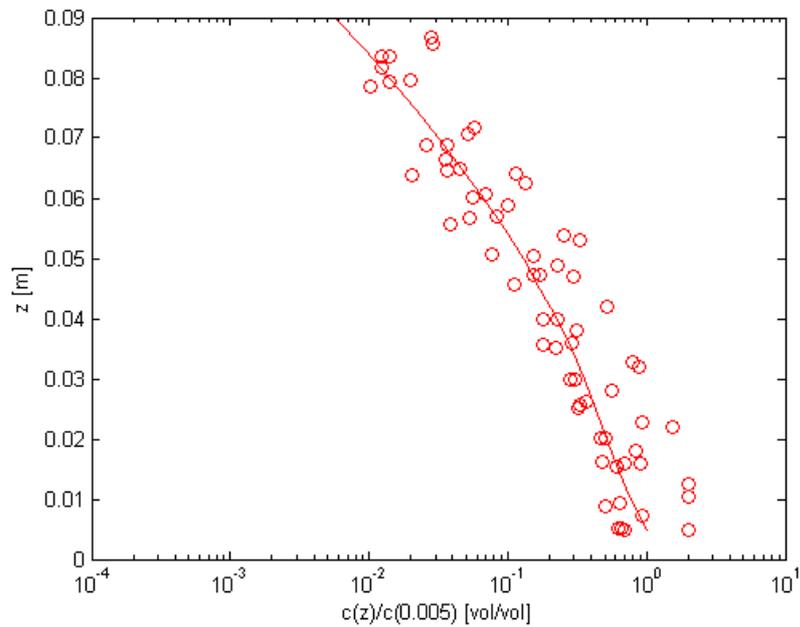

Fig. 7. Time-averaged concentration profile over wave ripples for fine sand ($w_s$=0.65cm/s). Curve: numerical solution to the third-order ODE (31) with $l_{m2}(z)=\lambda\left[L_w-(L_w-z_0)\exp\left(-\frac{(z-z_0)}{L_w}\right)\right]$ ; ($\lambda=1$) ; $L_w$=0.022m. O: measurements.

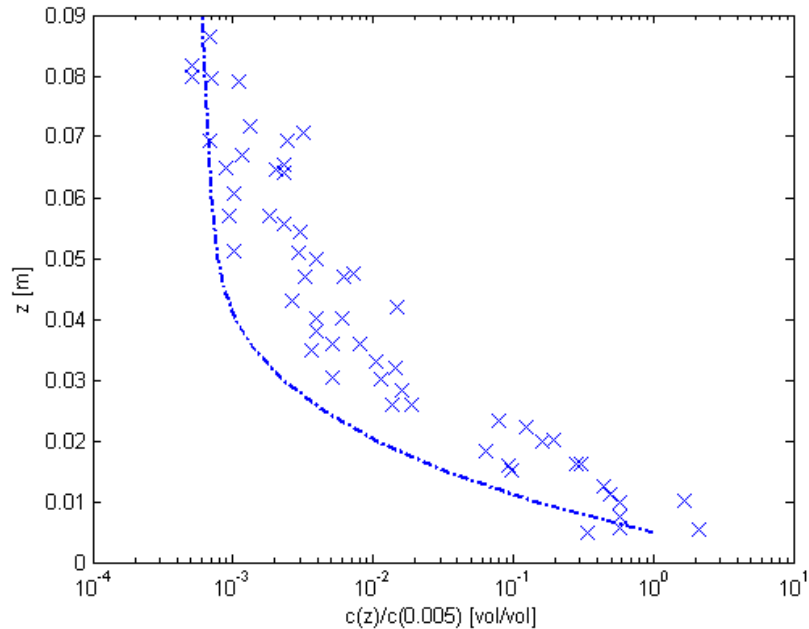

Fig. 8. Time-averaged concentration profile over wave ripples for coarse sand ($w_s$=6.1cm/s). Curve: numerical solution to the third-order ODE (31) with $l_{m1}(z) = \lambda z_0 \exp\left(3\frac{z-z_0}{L_w}\right)$ ; ($\lambda$=1) ;

$L_w$=0.05m , $z_0$=0.005m; $w_m(z_0)$=0.025m/s. X: measurements.

With $z_0$=0.005m; $w_m(z_0)$=0.025m/s and $L_w$=0.044m, figure (5) shows that $l_{m1}$ improves the solution for coarse sand, $l_{m1}$ increases the upward concavity of $C(z)$ (bold solid line for coarse sand) while figure (6) shows that $l_{m2}$ increases the upward convexity of $C(z)$ (bold dashed line for fine sand).

As we can see in figures (7) and (8), the fit could be improved by changing the value of $L_w$. The solution for fine sand needs a decrease in $L_w$, the curve in figure 7 corresponds to $l_{m1}$ with $L_w$=0.022m (Nielsen and Teakle's value), while the solution for coarse sand needs an increase in $L_w$, the curve in figure 8 corresponds to $l_{m2}$ with $L_w$=0.05m. These simulations show that for a same settling velocity, mixing length profiles change the shape of $C(z)$. It shows particularly the capacity of $l_{m1}$ to reproduce the shape of the measured concentrations for coarse sand (figure 8).

**CONCLUSION**

In this paper, we presented an analytical expression for the mixing velocity, and two theoretical mixing length equations. The two proposed mixing length equations are of different nature since the first mixing length profile is upward concave and increases with roughness while the second mixing length profile is upward convex and decreases with roughness. These differences are probably due to the related assumptions. The two mixing length equations are based on an exponential decrease for *k* and a local equilibrium between energy production and dissipation. In addition to these two assumptions, the second mixing length equation is based on the similarity hypothesis.

The proposed equations were used in the modeling of time-mean sediment concentration profiles in an oscillatory boundary layer over wave ripples. The first theoretical mixing length equation seems to be able to improve the prediction of the concentration profile for coarse sand.

Finally, the simulations presented here suggest that the $l_m$ profile varies with grain sizes. One final question remains. Could grain sizes change the shape of mixing length profile $l_m$? ($l_m$ concave downward for coarse sand and downward convex for fine sand). To be able to answer to this question, we need to carry out more investigations. Our two proposed $l_m$ equations and the related assumptions need more analyses.

**ACKNOWLEDGEMENTS**

The author would like to thank Peter Nielsen and Ian Teakle for providing experimental data.